\documentclass[reprint,aps,amsmath,amssymb,notitlepage,superscriptaddress, twocolumn,color,epsfig,graphicx,bm,floatfix,nofootinbib,longbibliography]{revtex4-1}
\pdfoutput=1
\usepackage[utf8]{inputenc}
\usepackage[english]{babel}
\usepackage[T1]{fontenc}
\usepackage{amsmath}
\usepackage{hyperref}

\usepackage{tikz}
\usepackage{lipsum}

\usepackage[utf8]{inputenc}

\usepackage{tikz}

\usetikzlibrary{decorations.markings}

\usepackage{amsmath}  
\usepackage{url}
\usepackage{hyperref}
\usepackage{mathtools}
\usepackage{siunitx}
\usepackage{multirow}
\usepackage{microtype}
\usepackage{array}
\usepackage{tabulary}
\usepackage{longtable}
\usepackage[version=4]{mhchem}
\usepackage{graphicx}
\newcolumntype{K}[1]{>{\centering\arraybackslash}p{#1}}

\usepackage{graphicx}   
\usepackage{verbatim}   
\usepackage{color}      
\usepackage{subfigure}  
\usepackage{hyperref}   
\newcommand{\btxt}[1]{{\color{black} #1}}

\usepackage{lipsum}

\begin{document}

\title{Demonstrating Quantum Scaling Advantage in Approximate Optimization for Energy Coalition Formation with 100+ Agents} 

 \author {Naeimeh Mohseni}

\affiliation {E.ON Digital Technology GmbH, Hannover, Germany}
\author {Thomas Morstyn}
\affiliation {Department of Engineering Science, University of Oxford, UK}
\author {Corey  O'Meara}
\affiliation {E.ON Digital Technology GmbH, Hannover, Germany}
\author {David Bucher}
\affiliation {Aqarios GmbH, Prinzregentenstraße 120, Munich, Germany}
\author {Jonas Nüßlein}
\affiliation {Ludwig-Maximilians University Munich, Institute for Computer Science, Germany}
\author {Giorgio Cortiana}
\affiliation {E.ON Digital Technology GmbH,  Hannover, Germany}

\begin{abstract}

The formation of energy communities is pivotal for advancing decentralized and sustainable energy management. Within this context, Coalition Structure Generation (CSG) emerges as a promising framework. The complexity of CSG grows rapidly with the number of agents, making classical solvers impractical for even moderate sizes. This suggests CSG as an ideal candidate for benchmarking quantum algorithms against classical ones.  Facing ongoing challenges in attaining computational quantum advantage for exact optimization, we pivot our focus to benchmarking quantum and classical solvers for approximate optimization. Approximate optimization is particularly critical for industrial use cases requiring real-time optimization, where finding high-quality solutions quickly is often more valuable than achieving exact solutions more slowly. Our findings indicate that quantum annealing (QA) on DWave can achieve solutions of comparable quality to our best classical solver, but with more favorable runtime scaling, showcasing an advantage.  This advantage is observed when compared to solvers, such as  Tabu search,  simulated annealing, and the state-of-the-art solver Gurobi in finding approximate solutions for energy community formation involving over 100 agents. DWave also surpasses 1-round QAOA on IBM hardware. 
Our findings represent the largest benchmark of quantum approximate optimizations for a real-world dense model beyond the hardware's native topology, where D-Wave demonstrates a scaling advantage.
\end{abstract}
\maketitle

\section{Introduction\label{sec:introduction}}

To explore the potential of quantum computing in optimization problems it is encouraging to benchmark quantum solvers against classical solvers in tackling problems that are sufficiently hard and exceed the capabilities of state-of-the-art classical solvers.  This exploration becomes even more intriguing when problems get already hard in intermediate sizes where existing quantum hardware can be effectively leveraged.  It is also crucial that the problem can be formulated compatible with existing quantum algorithms and hardware \cite{abbas2023quantum}. As we explain next, energy coalition formation applying the Coalition Structure Generation (CSG) game stands out as an exemplary candidate for such benchmarking, given its computational complexity and profound real-world significance within the industry. The formation of energy communities represents a significant step towards decentralized and sustainable energy management, offering opportunities for local empowerment, renewable energy adoption, and enhanced resilience in energy systems \cite{tushar2020coalition, luo2021multiple, blenninger2024quantum, blenninger2024quantum, nusslein2024towards, bucher2024evaluating}.

CSG provides a framework for understanding how agents,  can form coalitions denoted by $C_l$ (with $l$ being the index of the coalition) to optimize their collective benefits \cite{bandeiras2023application, moafi2023optimal, han2018incentivizing, rahwan2015coalition}. A coalition within this context can be characterized by a characteristic function, typically denoted as $v (C_l)$, which assigns values to different coalitions. 
The value function provides a measure of the cooperative advantage or payoff that the agents in the coalition can attain by working together, as opposed to operating independently \cite{bachrach2013optimal, deng1994complexity}.

The task of creating coalition structures is highly challenging, especially given the presence of $n$ agents, leading to a staggering $O(n^n)$ possible partitions. Currently, the only algorithm capable of finding an optimal solution 
is the Dynamic Programming algorithm, achieving this in $O(3^n)$ time \cite{rahwan2008improved}. This approach is limited to systems consisting of a few tens of agents ( i.e. $n<$ 25). CSG can be transformed into an approximately equivalent game represented using the induced subgraph game (ISG) \cite{9314263}. In this representation, nodes correspond to agents, and coalition values are reflected in pairwise interactions within a weighted graph, such that $v\left(C_l\right)=\sum_{(i, j) \in C_l} w_{i j}$ \cite{deng1994complexity, bachrach2013optimal}. The transformation of CSG to ISG allows the problem to be recast as quadratic unconstrained binary optimization (QUBO), enabling more computationally efficient approaches. Nonetheless, it is crucial to note that the problem remains NP-complete \cite{bachrach2013optimal} such that leading classical solvers, such as DyCE \cite{bistaffa2014anytime} and CFSS \cite{bistaffa2014anytime}  are limited in scalability and memory requirements. For instance, DyCE \cite{bistaffa2014anytime}, which employs dynamic programming, is restricted to systems with around 25 agents due to its exponential memory complexity ($O\left(2^n\right)$) and lacks an anytime approach.  CFSS can handle larger but shallow problems, yet still explores all possible solutions for complete graphs with a complexity of $O(n^n)$.

\begin{figure*}[t]
\centering
\includegraphics[width=1\linewidth]{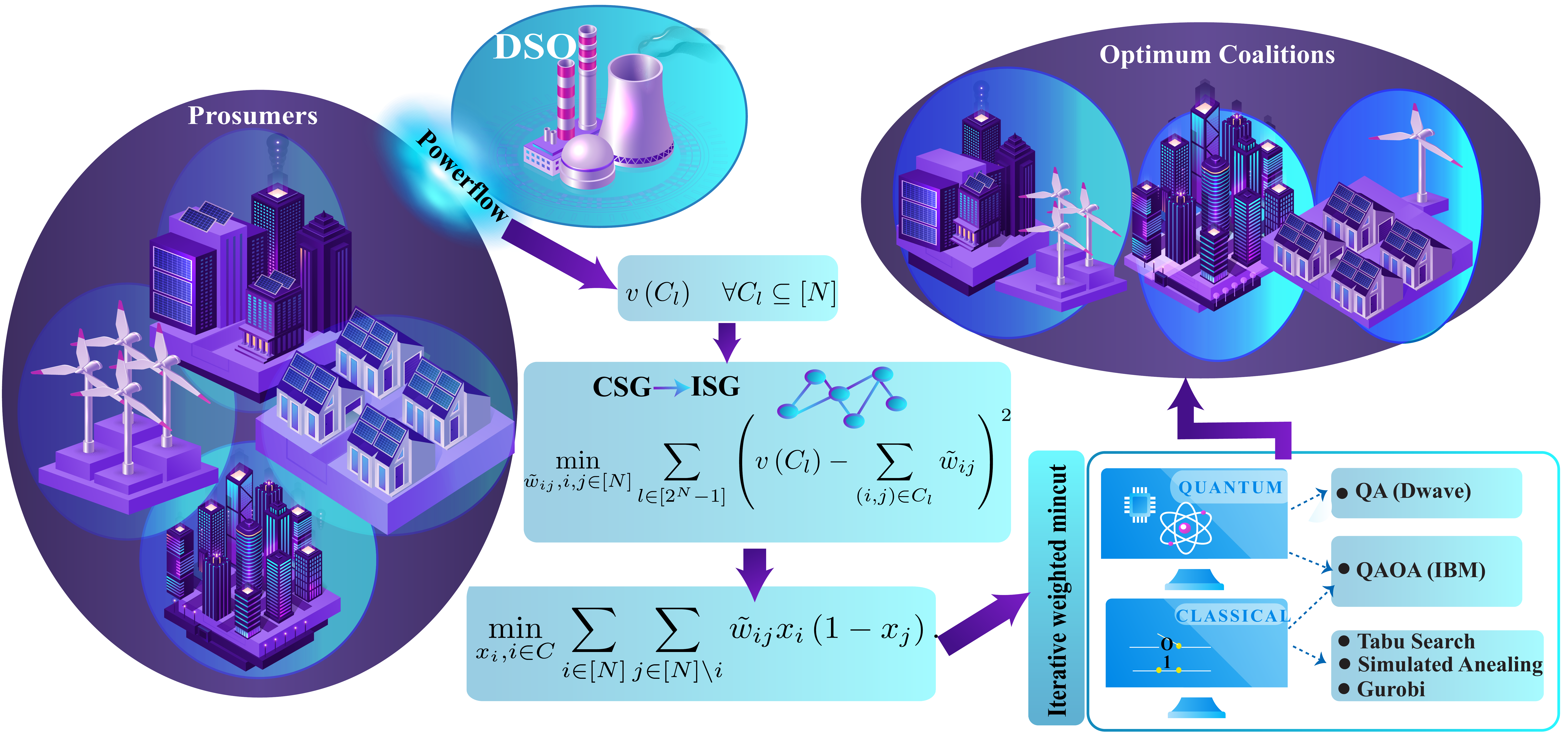} 
\caption {(a) Schematic representation of the workflow: The Distribution System Operator (DSO) strategically evaluates the potential of prosumer communities, aiming to minimize network management costs through optimized power flow analysis. The DSO's decision-making involves partitioning prosumers into net-metered coalitions, treating it as a  Coalition
Structure Generation (CSG) problem to optimize overall network efficiency. We transform the CSG formulation into the Induced Subgraph Game (ISG).  An approximate solution to the ISG is found through iterative splitting coalitions to bipartitions until no value-increasing bipartitions remain. In the benchmarking phase, the efficiency of both quantum and classical solvers is assessed in the iterative splitting process.
\label{fig1}}
\end{figure*}

 
In this work, we propose a novel formulation of energy coalition formation that can be encapsulated within the framework of CSG.
We first introduce a novel method for transforming the CSG formulation into an ISG.   Building on this transformation,  we apply the approximation introduced in \cite{venkatesh2023quacs}, under which our problem reduces to an iterative weighted minimum cut on the ISG (formulated as a QUBO). The complexity of finding an optimal solution for this reduced formulation remains $O(n^3)$. We show a solution where its optimality is not proven can be found in polynomial time.

Our performance evaluation involves benchmarking quantum annealing on DWave hardware and the Quantum Approximation Optimization Algorithm (QAOA)
on  IBMQ hardware against classical solvers such as  Gurobi \cite{gurobi}, Tabu search \cite{glover1986future}, simulated annealing \cite{kirkpatrick1983optimization}, and QB-solve  \cite{booth2017partitioning})  to find approximate solutions.  
\btxt{We provide preliminary empirical evidence suggesting a potential scaling advantage of DWave for a dense, real-world problem in \textit{approximate optimization}.}
A recent study \cite{PhysRevLett.134.160601} applying quantum annealing correction demonstrates a scaling advantage for approximate optimization against parallel tempering with isoenergetic cluster moves. However, that work is limited by the problem's compatibility with the hardware's connectivity and is focused on a 2D model, which restricts its range of applications. 
We also show Dwave outperforms 1-round QAOA on IBM Hardware.

Additionally, our experiments demonstrate that the solutions obtained through approximation on the ISG remain remarkably close to those derived from the original CSG formulation (up to system sizes that we have been able to check). This showcases the robustness and efficacy of our new introduced method in transferring CSG to ISG.

\section{Problem formulation \label{Sec:2}}

Consider a distribution network with $N$ prosumers (proactive consumers), each with some combination of inflexible loads, flexible loads, generation and/or storage. A standard arrangement for retail contracts is for each prosumer's energy demand to be individually metered, with the net demand during each metering period $t$ charged at an import price $\lambda_t^b$, and net supply rewarded at a lower export price $\lambda_t^s \leq \lambda_t^b$. An alternate arrangement is to enable groups of retail prosumers to form net-metered energy communities, where prosumers' collective net demand is metered. This creates an overall incentive for community supply-demand balancing, and cooperative game theory can be used to calculate cost/revenue sharing which aligns individual incentives with collective optimization. The energy management problem for each coalition with optimization horizon $[T]$, can be generically described by,



\begin{equation}
\begin{aligned}
\min_{p_i, i\in[C_l]} & \left( \sum_{i\in[C_l]} c_i(p_i) \right. \\
& \left. - \Delta_t \sum_{t\in[T]} \left( \lambda_t^b\left[\sum_{i\in[C_l]} p_{i t}\right]^- + \lambda_t^s\left[\sum_{i\in[C_l]} p_{i t}\right]^+ \right) \right) \\
& \text{s.t. } p_i \in \mathcal{P}_i, \forall i\in[C_l].
\end{aligned}
\end{equation}

 \btxt{$c_i\left(p_i\right)$  presents each prosumer's cost function associated with the usage of their flexible devices and a constraint set $\mathcal{P}_i$ for their output power vector, which may be time-coupled. For the exact choice of $c_i\left(p_i\right)$  and constraint set $\mathcal{P}_i$, refer to Supplemental Material Sec. II. We label the $N$ coalitions for which prosumers are metered individually $C^0_{i}$. We remind that $C_i$ refers to $i$-th colaition.  $p_i=\left(p_{i 1}, \ldots, p_{i T}\right)$ is the vector of output powers over the time horizon and $p_{i t}$ is the prosumer output power during time interval $t$.
 $\left[p_{i t}\right]^{+}=\max \left\{0, p_{i t}\right\}$ and $\left[p_{i t}\right]^{-}=\min \left\{0, p_{i t}\right\}$.    $\Delta_t$ is the duration of each time interval.}

The supply-demand balancing incentivized by community net metering is generally desirable for local groups of prosumers. However, when expanded to larger more dispersed groups, there is the potential that this will introduce distribution losses and the need for additional flexibility procurement to prevent network constraint violations.
Consider the perspective of a distribution system operator (DSO), and assume it has the authority to allow or disallow the formation of net-metered energy communities. The DSO needs to be able to calculate the value in terms of reduced network management costs (which could be negative) of allowing a coalition of prosumers $C_l $ to form a net-metered community. In general, network power flows and management costs will depend on the collective operation of all prosumers. To simplify the value calculation for coalition formation, we focus on the marginal value created assuming prosumers which are not in the coalition operate under individual metering. In this case, the value created can be approximated as,


\begin{equation}
\begin{aligned}
v\left(C_l\right) & =\mathcal{O}^{o p f *}\left(p_i^{\text {ind } *} \mid i \in[N]\right)-\mathcal{O}^{o p f *}\left(p_i^{\text {net } *} \mid i \in[N]\right), \\
& p_i^{\text {ind } *} =p_i^* \text { from (1) optimized for $C^0_i$, } \\
p_i^{\text {net* }} & = \begin{cases}p_i^* \text { from (1) optimized for $C_l$, } & \text { if } i \in C_l, \\
p_i^* \text { from (1) optimized for $C^0_{i}$, } & \text { otherwise }\end{cases}
\end{aligned}
\end{equation}
Here, ${ }^*$ indicates the solution to an optimization problem. $p_i^{\text {ind } *}$  ($p_i^{\text {net } *}$) represents the optimized power for the prosumer $i$ under individual (net) metering.  $[N]=\{1,2,...N\}$ and $\mathcal{O}^{o p f *}$ is the objective function value for the following optimal power flow problem which the DSO can use to procure flexibility to manage losses and network constraints,

\begin{equation}
\begin{aligned}
&\min_{\substack{p_{k t}^{\uparrow}, p_{k t}^{\downarrow}, k \in[K], t \in[T]}} \Delta_t \sum_{t \in[T]} \sum_{k \in[K]} \lambda_{k t}^{\uparrow} p_{k t}^{\uparrow} + \lambda_{k t}^{\downarrow} p_{k t}^{\downarrow} \\
\text{s.t.} \quad &\left\{p_{k t}^{\uparrow}, p_{k t}^{\downarrow} \mid k \in[K]\right\} \in \mathcal{N}_t\left(p_{i t}^* \mid i \in[N]\right), \forall t \in[T], \\
&\left\{p_{k t}^{\uparrow}, p_{k t}^{\downarrow} \mid t \in[T]\right\} \in \mathcal{P}_k^{\uparrow \downarrow}, \forall k \in[K].
\end{aligned}
\end{equation}

$p_{k t}^{\uparrow}, p_{k t}^{\downarrow}$ are the costs of upward/downward flexibility at network node $k \in[K]$ during time interval $t \in[T]$, and $\lambda_{k t}^{\uparrow}, \lambda_{k t}^{\downarrow}$ are the associated prices. $\mathcal{N}_t$ is the network constraint set for time interval $t$ (e.g. nodal power balance, bus voltage limits, line power flow limits). Note that this constraint set will depend on the prosumer output powers from the retail market $p_{i t}^*, i \in[N]. \mathcal{P}_k^{\uparrow \downarrow}$ is the constraint set for flexibility at network node $k$, which in general may be time-coupled. For more details on power flow optimization refer to Sec. II of supplemental Material.

The DSO decision problem is then to partition the prosumers into a set of net-metered coalitions with the goal of minimizing network management costs. This is formulated as a  CSG problem, which can be solved using binary integer linear programming (BILP) \cite{venkatesh2022bilp},
\begin{equation}
\begin{aligned}
\max _{x_l, l \in\left[2^N-1\right]} & \sum_{l \in\left[2^N-1\right]} v\left(C_l\right) x_l, \\
\text { s.t. } & \sum_{i \in\left[2^N-1\right]} S_{i l} x_l=1, \forall i \in[N], \\
& x_l \in\{0,1\}, \forall l \in\left[2^N-1\right] .
\end{aligned}
\end{equation}

Note that there are $2^N-1$ potential sub-coalitions of prosumers. $S_{i l}=1$ if prosumer $i \in[N]$ belongs to coalition $C_l$ (i.e. $i \in C_l$ ), and $S_{i l}=0$ otherwise. Binary decision variable $x_l=1$ indicates that $C_l$ is part of the selected partition, and the value function $v\left(C_l\right)$ gives the value of the coalition.
This formulation is computationally intensive since the number of binary decision variables depends on the number of sub-coalitions, which increases exponentially with the number of prosumers (i.e. $2^N-1$).

This CSG formulation can be transformed into ISGs, where the value of a coalition can be expressed as the sum of pair-wise ``joint utilities'' between prosumers $w_{i j} \in \mathbb{R}, i, j \in[N]$, i.e. $
v^{\text{ISG}}\left(C_l\right)=\sum_{(i, j) \in C_l} w_{i j}$ with lower computational requirements. The value function nonlinearities mean this framework is not directly applicable to our CSG. However, we can make use of an ISG based on pairwise joint utilities $\tilde{w}_{i j}$ which approximate our CSG. The following quadratic programming problem can be used to find pairwise weights that minimize the mean squared error between the value functions and an approximate ISG,
\begin{equation}
   \min _{\tilde{w}_{i j}, i, j \in[N]} \sum_{l \in\left[2^N-1\right]}\left(v\left(C_l\right)-\sum_{(i, j) \in C_l} \tilde{w}_{i j}\right)^2 . \label{Value_to_weight}
\end{equation}

Starting with the grand coalition, an approximate solution to the ISG can be found by iteratively splitting coalitions in two, until no value-increasing bipartitions are available \cite{venkatesh2023quacs}. \btxt{For a given iteration, let $C$, be one of the optimal bipartitions that was found for our $N$ prosumers. The optimal bipartitions of $C$ at the next iteration can be found using the following quadratic unconstrained binary optimization (QUBO) with $N$ binary decision variables $x_i$,}
\begin{equation}
\begin{aligned}
\min _{{x}_i, i \in C} & \sum_{i \in[N]} \sum_{j \in[N] \backslash i} \tilde{w}_{i j} x_i\left(1-x_j\right) . \\
\text { s.t. } & x_i \in\{0,1\}, \forall i \in[N] .
\end{aligned}
\end{equation}

\btxt{Here the binary vector $\mathbf{x}=\left[x_1, x_2, \ldots, x_N\right]$ for  $N$ prosumers encodes the bipartition: prosumers with the same value of $x_i$ belong to the same subcoalition. For example, $\mathbf{x}=[1,0,0,1]$ indicates that prosumers 1 and 4 form one subcoalition, while prosumers 2 and 3 form the other. }
 
 The schematic representation of the full workflow is shown in Fig. \ref{fig1}.


\section{Results}
\begin{figure*}[t]
\centering
\includegraphics[width=1\linewidth]{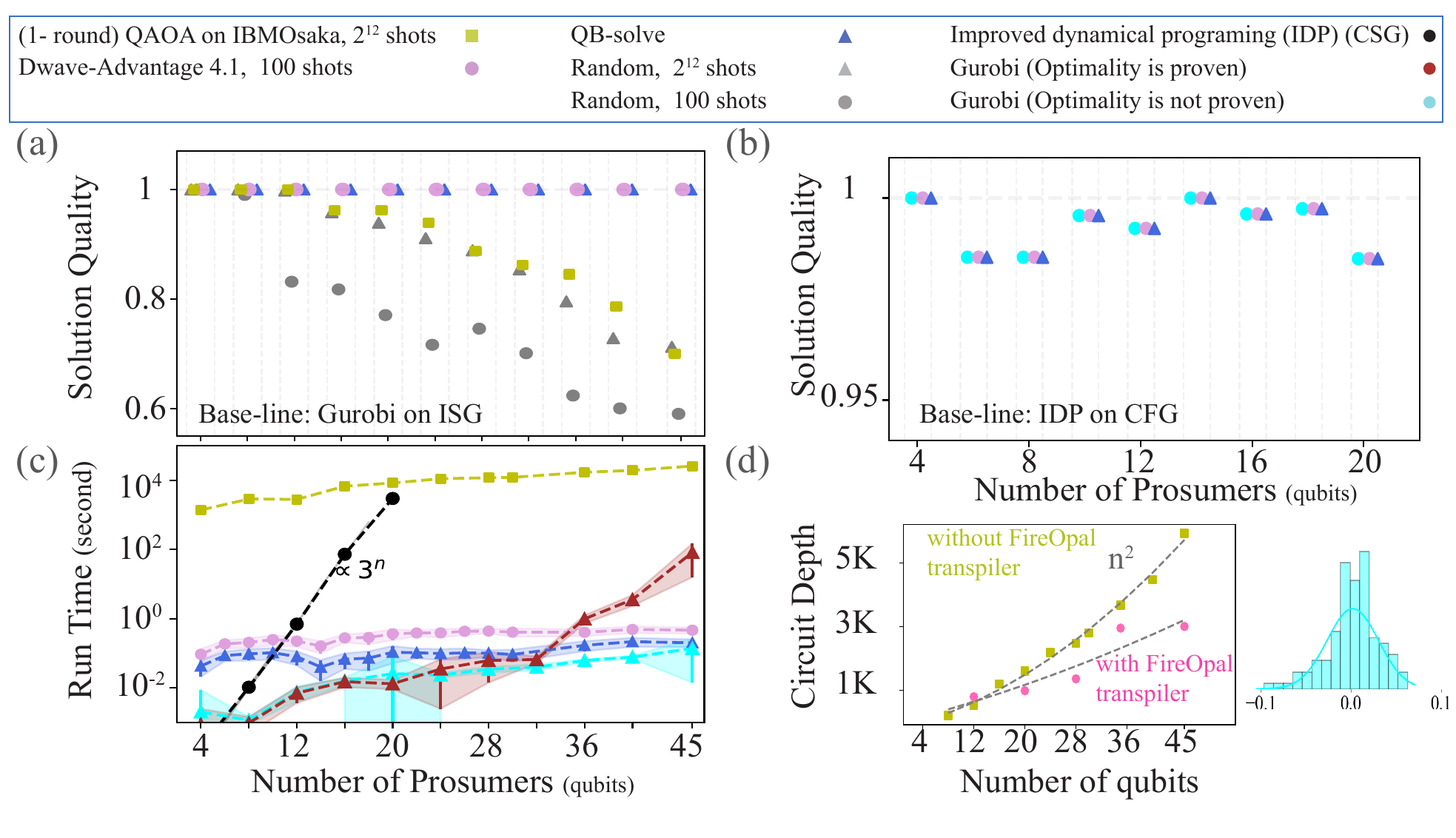} 
\caption {Benchmarking classical solvers against quantum solvers in energy coalition formation:   (a) Solution quality of different solvers on ISG formulation where the baseline is w.r.t. the solution determined with the Gurobi on the same formulation. (b) The solution quality of different solvers on ISG  formulation w.r.t. the solution determined with IDP on the original CSG formulation. (c) Run time of different solvers on ISG formulations as well as run time of improved dynamical programming (IDP) on CSG formulation.   (d) QAOA circuit depth on quantum hardware for different numbers of prosumers (number of qubits).  Note that results for each number of prosumers showcase averages across 20 realizations except for IBM hardware and random solver with $2^{12}$ shots where results are shown for a single problem instance per data point. The light blue histogram on the lower right shows the weight distribution after transferring CSG to ISG.
\label{fig2}}
\end{figure*}
In this section, we elaborate on the outcome of our performance evaluation, benchmarking an exact classical solver, the commercial state-of-the-art classical solver Gurobi, and QB-solve \cite{booth2017partitioning} against quantum
annealing on DWave hardware (DWave Advantage 4.1), and 1-round QAOA
on  IBMQ hardware (IBM-Osaka).    For a detailed overview of all the solvers utilized in our benchmarking,  refer to Supplemental Material Sec. I. 
 Before presenting the benchmarking results, it is essential to elucidate the structure and complexity of the problem instances generated from our use case as follows.

 \textbf{Problem complexity:} 
   Our analysis reveals that the graphs resulting from the mapping of CSG to ISG are very dense across all system sizes, with connectivity levels exceeding 90 percent. Moreover, the weight distribution includes both positive and negative values forming a Gaussian distribution centered closely around zero. The light blue histogram depicted in Fig. \ref{fig2} lower right panel illustrates the weight distribution for the case of 18 prosumers averaging over 20 instances (different instances of the problem correspond to different initial power configurations for prosumers at the outset and varying export and import prices). This scenario, characterized by a dense graph featuring a mixture of positive and negative weights, represents a particularly challenging case \cite{mccormick2003easy}. For further insights, refer to\cite{mccormick2003easy}, where they discuss the boundary between polynomially-solvable Max Cut and NP-Hard Max Cut instances when they are classified only on the basis of the sign pattern of the objective function coefficients.

\textbf{Solver compatibility:} When benchmarking algorithms, selecting the appropriate formulation for each solver is crucial to ensure their optimal performance. Our problem formulation here namely the QUBO formulation is well-suited to all the solvers we are applying in this work.

\textbf {Benchmark metrics:} 
We benchmark the performances in terms of solution quality and running time  (Fig. \ref{fig2}). The runtime for all solvers includes also any classical processing time, such as the time spent optimizing variational parameters. 
The results for each number of prosumers showcase averages across 20 realizations except for IBM hardware and random solver with $2^{12}$ shots where results are shown for a single problem instance per data point.

\textit{Solution quality:} In Fig. \ref{fig2}(a), we present the solution quality defined as the ratio of the optimal coalition values obtained by any solver on ISG formulation to the values found by Gurobi on the same formulation. Up to this size, Gurobi is still able to prove the optimality of the solutions found. For the prosumer sizes studied here (up to 45), both DWave and our classical solver consistently match Gurobi's solution.  In contrast, the 1-round QAOA exhibits a significant decline in solution quality as system size increases and its performance is indistinguishable from a random solver. In some cases, QAOA performs slightly better than a random sampler but mostly its performance is indistinguishable from a random solver. 

Here QAOA is enhanced with Fire Opal's error suppression pipeline, excluding any readout mitigation and postprocessing. This incorporates automated error-suppression techniques, such as layout optimization \cite{hartnett2024learning}, dynamical decoupling for crosstalk and dephasing suppression \cite{PhysRevApplied.20.064027}, and AI-driven gate waveform optimization.

Several factors contribute to the suboptimal performance of QAOA on 
 IBM hardware. The most significant one is that the traditional QAOA ansatz is inefficient and not hardware-friendly for dense problems like ours. In Fig. \ref{fig2} (d), we compare the circuit depth applying the Fire Opal transpiler and the default Qiskit transpiler, showing that circuit depth increases quadratically with the number of qubits using Qiskit's default. While Fire Opal’s transpilation significantly reduces the circuit depth and improves scaling, it still remains insufficient to make QAOA competitive for dense problems, given current gate fidelities \cite{matsuo2023sat, weidenfeller2022scaling}. Recent studies have proposed hardware-friendly modifications to the QAOA ansatz \cite{amaro2022filtering, PhysRevResearch.4.023249}, but their effectiveness for dense models remains elusive.

We also investigate the closeness of the solution derived from the ISG transformation to that derived from the original CSG formulation in Panel (b).  In this case, the solution quality is defined as the ratio of the optimal coalition value obtained from any solver on ISG to the exact solution of the original CSG formulation determined by improved dynamic programming (IDP) \cite{rahwan2008improved}. IDP is the only existing algorithm capable of finding an optimal solution for the CSG formulation. However, the runtime scales as 
$O(n^3)$, allowing us to calculate the solution quality for up to 20 prosumers. It is very encouraging to see that solutions obtained through approximation on the ISG
remain remarkably close to those derived from the original
CSG formulation. This confirms the efficacy of the mapping that we introduced in Eq. \ref{Value_to_weight}.

\textit{Run Time:} Fig. \ref{fig2} (c), shows the run time of different solvers.
DWave and QB-Solve demonstrate competitive runtimes, but clear scaling is difficult to observe with up to 45 prosumers. For Gurobi, we report two runtimes: the cyan line represents the time taken to find a solution that no longer improves over time, though its optimality is not proven. The brown line shows the total runtime, including the time spent proving the solution's optimality. As the number of prosumers increases, it becomes increasingly challenging for Gurobi to confirm optimality. For instances with more than 50 prosumers, Gurobi is unable to prove optimality even after several hours of computation.  IDP on CSG also experiences a noticeable increase in runtime as the problem size grows.  

Interestingly,  the scaling behavior of QAOA on Hardware is close to DWave and QB-solve, albeit with a significantly larger prefactor. This can be attributed to iteratively updating variational parameters and running the algorithm on hardware for each parameter update. Initializing variational parameters based on existing heuristic schemes \cite{PhysRevX.10.021067, akshay2021parameter} or learning them via machine learning techniques \cite{ montanez2024transfer, mohseni2023deep, mohseni2023deepgap} could potentially reduce this prefactor to some extent.

To compare our solvers effectively, it is essential to consider both runtime and solution quality together. For problem sizes up to 45 prosumers, both D-Wave and our classical solver consistently achieve matching solutions; however, runtime comparisons at this scale do not provide reliable insights into scaling behavior. Our problem instances become difficult to solve beyond 50 prosumers, where even Gurobi struggles to find optimal solutions. Therefore, drawing firm conclusions about performance requires exploring larger instances, which we address next.

\begin{figure*}[th]
\centering
\includegraphics[width=1\linewidth]{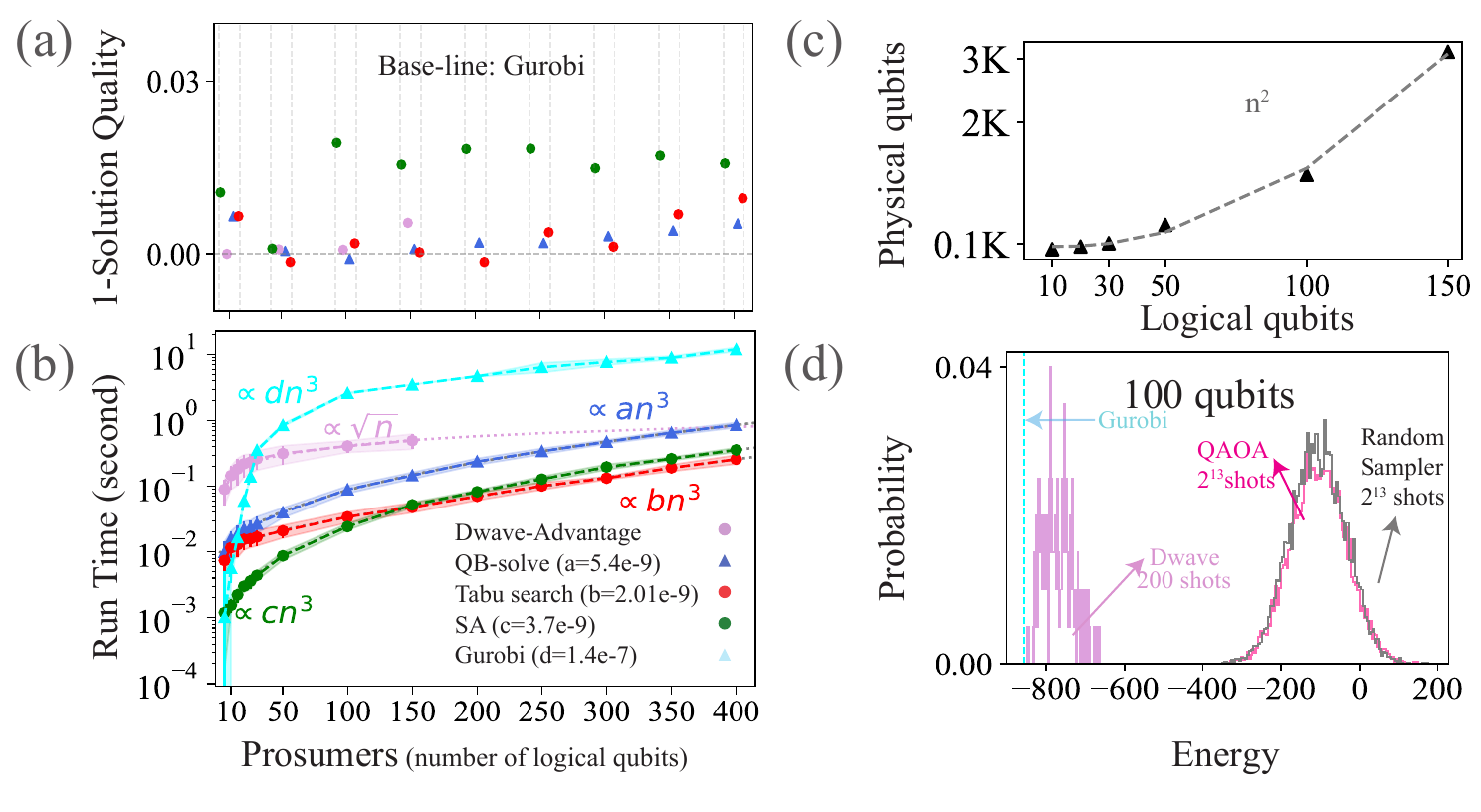} 
\caption {Benchmarking classical solvers against DWave in energy coalition formation for randomly generated problem instances.  (a) The solution quality versus the number of prosumers. (b) Run time versus number of prosumers.   (c) The number of physical qubits versus logical qubits (number of prosumers) on DWave hardware for the first split of the grand coalition into two partitions.  Note that results for each number of prosumers showcase averages across 20 realizations. (d) Energy distribution sampled from DWave, QAOA,  and a random sampler for the first split of the grand coalition to 
 two partitions for a randomly generated instance with 100 prosumers (100 qubits). The dashed vertical line shows the solution found by Gurobi.
\label{fig3}}
\end{figure*}

 \textbf{Randomly generated problems:}
To enhance the robustness of our findings and gain deeper insights into the performance scalability of DWave, we validate our results on larger-scale random problems by generating our problem instances randomly. Solving power flow equations for more than a few tens of prosumers (required to generate problem instances) demands substantial computational resources, and for the scope of this work, random generation suffices. For smaller numbers of prosumers, the distribution of generated problems, after transforming from CSG to ISG, follows a Gaussian distribution centered around zero with a width of 0.2 across all prosumer sizes. Therefore, for larger sizes \btxt{(above 22 prosumers)}, we generate our problem instances using the same Gaussian distribution but intentionally vary the width to 2. This deliberate variation allows us to investigate the correlation between solvers' performance and distribution characteristics, which may align with different choices of constraints and parameters in the problem formulation

The outcome of this investigation is presented in Fig. \ref{fig3}. For a more comprehensive benchmark, we additionally employ two other classical solvers: simulated annealing and Tabu search.  
\btxt{For instances with more than 50 prosumers, Gurobi fails to find a provably optimal solution and fails to terminate even after several hours of computation.} Additionally, we observed that the objective value no longer improves significantly after some time. Therefore, we introduced a termination criterion: the optimization process is stopped when the change in the objective value is less than 1e-8, and no significant improvement is observed within a predefined time window of 1 second. Note that if no improvement occurs within this 1-second window, the time is not included in the reported runtime. We consider the best solution found by Gurobi as the reference, even though its optimality remains unproven and other solvers may find a better solution.  

The optimality gap—the difference between the best solution found by the solver and the upper bound of the optimal solution, derived from the dual problem—varies across different problem instances and iterations, particularly during the process of splitting coalitions into bipartite groups. We report the optimality gap for the first iteration, namely the initial split of the grand coalition into two partitions. Averaged across all problem instances, the gap is $17\%, 26\%, 25\%, 29\%, 28\%, 30\%, 29\%$ for prosumer sizes of 50, 100, 150, 200, 250, 300 and 350, respectively. In most cases, the optimality gap reaches zero in later iterations. 

DWave consistently demonstrates performance in line with the results depicted in Fig. \ref{fig2} and competes effectively with the classical solvers.
 In terms of solution quality, for problem sizes up to 150 prosumers, the largest scale tested on the hardware, D-Wave consistently finds solutions similar to those obtained by Gurobi, except for the case with 150 prosumers, where D-Wave's solution is $1\%$ suboptimal. Within this range of prosumers, Tabu search consistently finds solutions that are either slightly better or worse than those found by Gurobi, within less than $1\%$ difference. 
For instances with over 200 prosumers, where no results are available from D-Wave, Gurobi consistently identifies the best solution among classical solvers, and all solvers outperform SA.

In terms of run time, DWave demonstrates a more favorable runtime scaling as a function of problem size, as depicted in Fig. \ref{fig3}(b).  The runtime of Gurobi, QB-solve, Tabu search, and simulated annealing scales cubically with varying prefactors, with Tabu search exhibiting the lowest prefactor. In contrast, DWave's runtime scales as $\sqrt n$ with $n$ the number of prosumers.

When using the DWave quantum computer to solve a problem, the connectivity of the logical qubits in the problem must be mapped onto the hardware's connectivity graph, which, for Advantage quantum processing units, is based on the Pegasus graph topology. This mapping process typically necessitates additional qubits, known as physical qubits, beyond those required by the original problem. Figure \ref{fig3} (c) illustrates the scaling of physical qubits with the number of logical qubits (prosumers). 

 In Fig. \ref{fig3} (d), we present the energy distribution (for a randomly generated instance with 100 prosumers) sampled from QAOA enhanced with Fire Opal’s error suppression pipeline \cite{sachdeva2024quantum}, DWave, and a random sampler. These distributions correspond to the first split of the grand coalition into two partitions. The vertical cyan line marks the solution found by Gurobi, though its optimality is unproven. Notably, DWave recovers the same solution, while the QAOA distribution closely resembles that of the random sampler. The circuit depth for  QAOA in this experiment applying the Fireopal's transpiler is 11k. Note that no postprocessing is applied to the QAOA output.

 \section{Conclusion and outlook\label{conclusion}}
The formation of energy communities signifies a pivotal shift towards decentralized and sustainable energy management.  In this work,  we propose a novel formulation of energy
coalition formation that can be encapsulated within the
framework of Coalition Structure Generation to comprehend how prosumers can unite in coalitions to maximize their collective advantages.  We also introduce a novel method for
transforming this formulation into a QUBO formulation for quantum solvers. 
We then conduct a benchmark comparing classical solvers with quantum 
solvers in this problem domain. Our findings showcase that the DWave machine outperforms 1-round QAOA on IBM hardware in terms of solution quality. 
\btxt{D-Wave exhibits more favorable time scaling with problem size than classical approximate solvers. However, for the system sizes we evaluated, it remains slower than some classical methods, such as QB-Solve. In terms of solution quality, D-Wave matches our best classical solvers (e.g., Gurobi and Tabu Search) for up to 100 prosumers (qubits) and achieves solutions within a $1\%$ optimality gap from Gurobi for problems with 150 prosumers (150 logical qubits)}.  

\begin{acknowledgments}
 N.M. and C.O. would like to thank Supreeth Mysore, Ivan Angelov, Stefan Wörner, Antonio Macaluso,  Gabriele Agliardi, and Rowen Wu for fruitful discussions.
 T. M. acknowledges the UK Engineering and Physical Sciences Research Council (EPSRC) (grant reference number EP/Y004418/1).

This work was supported by the German Federal Ministry of Education and Research under the funding program ``Förderprogramm
Quantentechnologien – von den Grundlagen zum Markt'' (funding program quantum technologies — from basic research to market), project
Q-Grid, 13N16177.
\end{acknowledgments}


\setcounter{equation}{0}
\setcounter{figure}{0}
\setcounter{table}{0}
\setcounter{section}{0}
\setcounter{subsection}{0}

\makeatletter
\renewcommand{\theequation}{S\arabic{equation}}
\renewcommand{\thefigure}{S\arabic{figure}}

\begin{widetext}

   \begin{center}
\textbf{\large Supplemental Material}
	\end{center}

	In this Supplemental Material, we provide a brief review of the algorithms that we applied in this work. We also provide further details on the formulation of our energy community formation problem.
\end{widetext}

\maketitle

\section{Algorithms}

In this section, we provide a concise overview of the operational mechanisms of  classical and quantum algorithms employed in this study.

\subsection{Quantum Algorithms}
\subsubsection{Quantum Annealing } 
\begin{figure}[ht]
\centering
\includegraphics[width=1\linewidth]{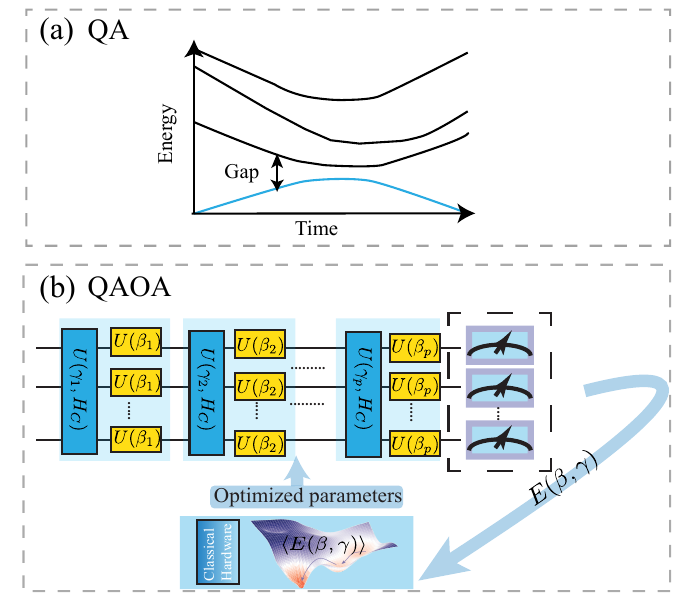} 
\caption {Schematic representation of QA and QAOA. \label{fig1s}}
\end{figure}

Quantum annealing (QA) is a heuristic algorithm based on the quantum adiabatic theorem  \cite{farhi2000quantum, RevModPhys.90.015002}.  In this algorithm, the system is initially prepared in the ground state of a Hamiltonian $H_0$ where its ground state is known.  A common choice for this initial Hamiltonian is $H_{0}=-\sum_{i=1}^{N} \sigma_{i}^{x}$, where $N$ denotes the number of qubits, and the ground state is the uniform superposition of all possible configurations $|+\rangle^{\otimes N}$ with $|+\rangle=(|0\rangle+|1\rangle)/\sqrt{2}$. The Hamiltonian is  gradually reweighted to the desired problem Hamiltonian $H_C$ according to
\begin{align}
H =(1- \lambda (t)) H_0+ \lambda (t) H_C ,
\label{Isingham}
\end{align} 
where $\lambda (t) \in [0,1] $ is the annealing schedule. 
The annealing process can be viewed as $H_0$ introducing quantum fluctuations originating from the non-commutability of  $H_C$ and $H_0$.  These fluctuations are gradually reduced to reach the low-energy configuration of the classical energy function $H_C$. Based on the quantum adiabatic theorem, if one performs the sweep sufficiently slowly, the system remains in its instantaneous ground state throughout the evolution (Fig. \ref{fig1s} (a) ) \cite{farhi2000quantum, RevModPhys.90.015002}. $H_C$ in our formulation is Eq. (6) at each min cut iteration. The number of qubits is then the number of prosumers. The use of quantum fluctuations in QA has been hypothesized as a potential resource for a speedup over classical methods. Quantum tunneling allows the system to pass through energy barriers that have a higher energy than available in the state \cite{denchev2016computational}.

\subsubsection{Quantum Approximate Optimization Algorithm}
QAOA can be seen as a troterrized version of QA. In QAOA \cite{farhi2014quantum}, the system is initially prepared in  $|+\rangle^{\otimes N}$.  Then, unitary operators $U(\gamma_{p},H_C) = e^{-i \gamma_{p} H_{\rm C}}$ (corresponding to the cost Hamiltonian $H_c$) and $U(\beta_{p}) = e^{-i \beta_{p}H_0}$ (correspond to the  mixer Hamiltonian $H_0=\sum_{i=1}^{N} \sigma_{i}^{x}$ are applied, alternatively, generating the following variational quantum state:
\begin{equation}
     |\psi(\boldsymbol{\beta}, \boldsymbol{\gamma})\rangle=e^{-i \beta_{\rm p} H_{0}} e^{-i \gamma_{\rm p} H_{\rm C}} \cdots e^{-i \beta_{1} H_{0}} e^{-i \gamma_{1} H_{\rm C}}|+\rangle^{\otimes N} ,
 \end{equation}
where $ \boldsymbol{\gamma}=\left(\gamma_{1}, \gamma_{2}, \cdots, \gamma_{\rm p}\right) \in[0, 2\pi]^p$ and $\boldsymbol{\beta}=\left(\beta_{1}, \beta_{2}, \cdots, \beta_{\rm p}\right) \in[0, \pi]^p$ are $2p$ variational parameters and $p$ determines the QAOA depth (number of rounds). Then, a classical optimizer is applied to find the optimal $( \boldsymbol{\beta}, \boldsymbol{\gamma}) $ that optimizes the energy expectation $
    E(\boldsymbol{\beta}, \boldsymbol{\gamma})=\left\langle\psi(\boldsymbol{\beta}, \boldsymbol{\gamma})\left|H_{\rm C}\right| \psi(\boldsymbol{\beta}, \boldsymbol{\gamma})\right\rangle $
by updating the variational parameters iteratively (In this study, we applied Cobyla as our classical Optimizer). The parameterized quantum circuit transfers the initial state to the ground state of the target problem Hamiltonian. Fig. \ref{fig1s} (b) shows the schematic representation of the algorithm.

\subsection{Classical Algorithms}
\subsubsection{Tabu Search}

Tabu search is a metaheuristic optimization algorithm  \cite{palubeckis2004multistart, glover1986future}.  Inspired by the concept of ``taboo'' this method strategically avoids revisiting previously explored solutions, preventing the algorithm from becoming trapped in local optima. By employing a short-term memory mechanism, Tabu search navigates the solution space by iteratively exploring neighboring solutions while adhering to tabu criteria that guide the search toward promising regions. Its balance between exploration and exploitation makes it a versatile and effective tool for addressing optimization challenges.
\subsubsection{QB-solve}
 The QB-solve tackles the challenge of finding the minimum value of large quadratic unconstrained binary optimization (QUBO) problems by breaking them into manageable pieces. It decomposes the global problem into smaller sub-problems by fixing certain bits and optimizing for those with the largest energy impact. This decomposition strategy allows QBsolv to focus on specific regions of the solution space where improvements are likely to have the greatest impact. Each piece is then solved using a specified solver, where here we use the Tabu algorithm \cite{booth2017partitioning}.
 \subsubsection{Simulated annealing}

Simulated annealing \cite{kirkpatrick1983optimization}, inspired by the physical process of annealing in metallurgy, stands as a heuristic tool in the realm of optimization algorithms. At its core, simulated annealing begins with an initial solution and iteratively explores nearby solutions, accepting those that improve upon the current state while occasionally allowing for worse solutions to be accepted probabilistically. This acceptance of inferior solutions is controlled by a temperature parameter, which acts as a guiding force: at higher temperatures, the algorithm is more likely to accept poorer solutions, akin to the exploratory phase of the annealing process. However, as the algorithm progresses, this temperature parameter gradually decreases, reducing the likelihood of accepting suboptimal solutions and guiding the search towards convergence.
\subsubsection{Gurobi} Gurobi \cite{gurobi}  Gurobi is a state-of-the-art commercial solver for a wide range of optimization problems, including linear programming, mixed-integer programming, and quadratic programming. It uses advanced techniques like branch-and-cut, combining the cutting plane method to tighten the solution space and the branch-and-bound algorithm to systematically explore possible solutions.  In Gurobi, the optimality gap is defined as the difference between the best solution found by the solver, and the best upper bound estimation of the optimal solution, estimated through the dual problem. A zero gap implies that Gurobi's solution is exact.
 \subsubsection{Random Sampler}
In this work, we apply the random sampler in the dimod package \footnote{\href{https://docs.ocean.dwavesys.com/en/stable/docs_dimod/}{https://docs.ocean.dwavesys.com/en/stable/docsdimod/}}. The random sampler generates a specified number of random configurations (number of shots specified by the user), which represent assignments of values to variables within the problem space defined by the binary quadratic model (BQM). Each configuration serves as a potential solution to the optimization problem. Subsequently, the sampler calculates the objective value for each generated configuration. Among these configurations, the sampler identifies and selects the one with the lowest objective value as the solution.

\section{Technical details on problem formulation}
The cost function $c_i\left(p_i\right)$ in Eq (1) can be chosen to be either linear or nonlinear, depending on the specific characteristics of prosumers and the energy system. We opt for a linear cost function, $c_i\left(p_i\right)=a_i p_i+b_i$, with $a_i=1$ and $b_i=0$ for simplicity. This choice may not fully capture all nuances of the prosumer's behavior and system dynamics. However, it suffices for our purpose.

We establish distinct constraints for agents based on their roles as prosumers, pure consumers, or pure producers at each time. We designate 90 percent of all agents as prosumers, denoting individuals capable of both producing and consuming energy. The remaining 10 percent are randomly assigned as either pure consumers or pure producers. 
For all prosumers across all time steps, constraints on power levels ($p_{it}$) in Eq. (1) are established relative to their initial power levels ($p^{\text{initial}}_{it}$).     This allows for fluctuations within a defined flexibility $\epsilon$, set to 0.5 here.
The initial power of prosumers are sourced from a simulation package that selects a specific location (e.g. Munich) and gathers information, including the number of residential, commercial, and industrial buildings. This data, leveraging standard load profiles and incorporating factors such as the number of households, shops, and industries, as well as their energy usage patterns, enables the estimation of power over time.

\begin{figure}[t]
\centering
\includegraphics[width=1\linewidth]{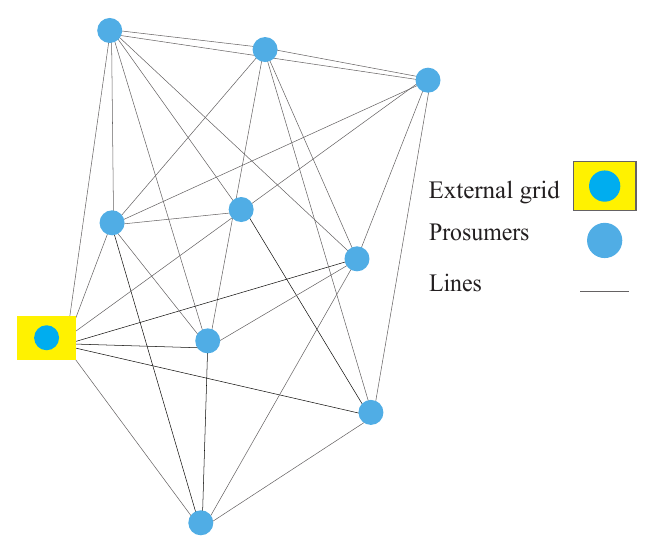} 
\caption {Schematic representation of the network with 9 prosumers and one external grid. \label{fig2s}}
\end{figure}
For agents that are prosumers, the power is bounded as follows

\begin{itemize}
    \item When  ($p^{\text{initial}}_{it} > 0$):
    \[ -p^{\text{initial}}_{it}(1 + \epsilon) \leq p_{i,t} \leq p^{\text{initial}}_{it}(1 + \epsilon) \]
    
    \item When ($p^{\text{initial}}_{it} < 0$):
    \[  p^{\text{initial}}_{it}(1 + \epsilon) \leq p_{i,t} \leq -p^{\text{initial}}_{it}(1 + \epsilon) \]
\end{itemize}

For agents that are pure producers ($p^{\text{initial}}_{it} > 0$) the power is bounded as:

    \[ 0 \leq p_{i,t} \leq p^{\text{initial}}_{it}(1 + \epsilon) \]

For agents that are pure consumers ($p^{\text{initial}}_{it} < 0$) the power is bounded as:

    \[ p^{\text{initial}}_{it}(1 + \epsilon) \leq p_{i,t} \leq 0 \]

The import and export price schedules are established by analyzing the $CO_2$ intensity derived from real-world data collected in Munich \footnote{\href{https://www.bayernwerk.de/de/fuer-zuhause/oekoheld.html}{https://www.bayernwerk.de/de/fuer-zuhause/oekoheld.html}}. In our study, we utilize an external grid, which functions as a Distribution System Operator (DSO). The external grid acts as a point of connection for exchanging energy to/from prosumers. A schematic representation of our network is shown in Fig. \ref{fig2s}. We consider four time steps.  Each time step represents a 15-minute interval, capturing a comprehensive one-hour time window.

For power flow optimization, we use Panda Power, a Python-based library for power systems\cite{thurner2018pandapower}. We employ the DC optimal power flow (DCPF) method within the Panda Power framework to optimize power flow across the network. 
Within Panda Power, we model the cost behavior using piece-wise linear (PWL) cost functions. These functions provide a realistic representation of generation costs by capturing nonlinear cost variations.

%
\providecommand{\noopsort}[1]{}\providecommand{\singleletter}[1]{#1}%

\end{document}